\documentclass[a4paper,twocolumn]{autart}

\usepackage{times,bm}

\usepackage{graphicx,psfrag,enumerate,subfigure}
\usepackage{amssymb}
\usepackage{amsmath}
\usepackage{calrsfs}
\usepackage{subeqnarray}
\usepackage{latexsym}

\newtheorem{definition}{Definition}
\newtheorem{theorem}{Theorem}
\newtheorem{remark}{Remark}

\newtheorem{lemma}{Lemma} 
\newtheorem{corollary}{Corollary}

\begin{document}
\begin{frontmatter}

\title{Gain-scheduled synchronization of parameter
  varying systems via relative $H_\infty$ consensus with  application
 to synchronization of uncertain bilinear systems\thanksref{footnoteinfo}}
                                % Title, 
                               % preferably not more than 10 
                                % words. 

\thanks[footnoteinfo]{This work was supported under Australian Research
  Council's Discovery Projects funding scheme (projects DP0987369 and
  DP120102152).} 

\author[First]{V.~Ugrinovskii}\ead{v.ugrinovskii@gmail.com} 

\address[First]{School of Engineering and IT, University of New South Wales
  Canberra, Canberra, ACT, 2600, Australia}


\begin{abstract} 
 The paper considers a problem of consensus-based synchronization of
 uncertain parameter varying multi-agent systems. We present a method for  
 constructing consensus-based synchronization protocol
 schedules for each agent to ensure it synchronizes with a reference
 parameter-varying system. The proposed protocol guarantees a specified
 level of $H_\infty$ transient 
 consensus between the agents. The algorithm uses a
 consensus-preserving interpolation and produces continuous (in the
 scheduling parameter) coefficients for the protocol. An application
 to synchronization of uncertain bilinear systems is discussed. 
\end{abstract}

\end{frontmatter}

%%%%%%%%%%%%%%%%%%%%%%%%%%%%%%%%%%%%%%%%%%%%%%%%%%%%%%%%%%%%%%%%%%%%%%%%%%

\section{Introduction}

The problem of synchronization of complex dynamical networks of
interconnected systems has received much
attention recently in the context of multi-agent
consensus~\cite{LDCH-2010,OFM-2007,YSS-2011,GYSS-2012}. In particular,
robust synchronisation of uncertain systems 
has gained attention recently~\cite{LDCH-2010,U6,TTM-2013}, including the 
development of $H_\infty$ methodology for multi-agent consensus and
synchronization. 
For instance, \cite{U6} considered an $H_\infty$ consensus
synthesis of observer-based synchronization protocols for dynamical
networks of uncertain agents and introduced the notion of
$H_\infty$ disagreement between agents to quantify consensus performance of
such networks. The model in~\cite{U6} allows for a
significant heterogeneity --- not only the agents are subject to additive
perturbations but they are also endowed with different sensing patterns. 

The $H_\infty$ metric has been used as measure of robustness in a
number of recent papers on multi-agent consensus and synchronization. The key
distinction of the $H_\infty$ consensus-based approach of \cite{U6} (also
see~\cite{LaU1,U7}), which is also used in this paper, is its focus on
disturbance attenuation to improve transient consensus. It was observed
in~\cite{U6} that $H_\infty$ optimization of consensus transients forces
consensus objectives to become a priority for the agents, thus creating
an essential prerequisite for synchronization. 

This paper extends the approach of $H_\infty$ 
consensus performance proposed in \cite{U6} to  
systems which require a 
time-varying reference for synchronization. Time variations of
systems coefficients pose an additional difficulty in addressing system
robustness, since many standard robust control 
and filtering techniques developed for time-invariant systems are not
directly transferable to time-varying systems. While in some situations the
issue can be circumvented by 
restricting attention to a finite-horizon version of the
problem~\cite{SWH-2010}, such an approach may not be suitable in
synchronization problems. For 
instance, synchronization of nonlinear systems exhibiting chaotic
behaviour requires the system to be continuously `locked' into
synchronous operation, otherwise even small discrepancies between
trajectories will cause the system to lose synchrony in a very short
time. 

Gain-scheduling techniques provide an alternative to the
finite-horizon analysis and synthesis of time-varying
systems 
known as parameter-varying systems. Such systems, especially their linear
versions termed linear parameter varying (LPV) systems, frequently arise in
the control systems theory~\cite{Shamma-Athans-TAC}.  
The 
controller design for such systems involves scheduling a
number of controllers for different operating conditions. 
When the operating conditions are continuously monitored, a
gain-scheduled controller can be designed by 
interpolating 
these controllers. This
allows to avoid detrimental transients caused by controller switching. Stability and robustness
preserving interpolation techniques for gain 
scheduling~\cite{Stilwell-Rugh-2000,Stoustrup-Komareji,YUMP1} provide a 
guarantee that the system governed by an interpolated 
controller remains stable  while traversing between operating points. 
 
Switching is particularly undesirable when the system employs high
gains. Since the distributed
estimation techniques in \cite{U6} were found to yield high gain
observers in some cases, in this paper we develop the interpolation
technique for gain-scheduled 
synchronization
inspired by the stability and
robustness preserving interpolation methods~\cite{Stilwell-Rugh-99,Stilwell-Rugh-2000,YUMP1}.    
The proposed interpolated protocol preserves
synchronization and 
$H_\infty$ consensus properties of the interpolants. 

As is well known large variations in the system parameters usually lead to a
conservative uncertainty 
model, resulting in infeasible $H_\infty$ design conditions or yielding
controllers with unacceptably poor performance; the approach of $H_\infty$
disagreement optimization in~\cite{U6} is not immune to this. The
scheduling method proposed in this paper can alleviate this
conservatism since it allows to consider multiple design points with
smaller parameter variations at each point. This is a certain advantage of
the proposed method over~\cite{U6}.

The paper further expands the preliminary results announced in \cite{U8a}
and applies them to the problem of 
synchronization for networks of uncertain single-input bilinear agents. In this
problem, the 
bilinear `master' system is assumed to be governed by a control signal
known to each agent. Such a signal can be generated by the reference system
and then made available to each agent, e.g., as  in the
`master-slave' synchronization
scheme~\cite{NM-1997,PC-1990}, or can be 
collectively generated by the network. 
We show that this signal can be used for protocol scheduling for each agent.
As an illustration, 
synchronization of a network of
chaotic oscillators is discussed.  

The paper is organized as follows. In Section~\ref{Formulation} we
introduce the problem, first for a network of bilinear systems, and then
for a more general class of parameter varying  uncertain systems which
include a globally Lipschitz nonlinearity. The main results of the paper
are developed in Section~\ref{main.results}, and the application of these
results to synchronization of single-input bilinear systems is given in
Section~\ref{sec.bi}.  The illustrating example is given in
Section~\ref{example}, followed by the Conclusions.  The proofs of the
results are collected in the Appendix. 

\paragraph*{Notation}  $\mathbf{R}^n$ denotes the real
Euclidean $n$-dimensional vector space, with the norm $\|x\|\triangleq
(x'x)^{1/2}$; the symbol $'$ denotes the transpose of a matrix or a
vector, and $\otimes$ is the Kronecker product of two matrices. 
Given a symmetric $k\times k$ matrix $P$, 
$\lambda_{\min}(P)$ denotes the smallest eigenvalue of $P$. 
The symbol $\star$ in  position $(k,l)$
of a block-partitioned matrix denotes the transpose of the $(l,k)$ block of
the matrix. $\mathbf{1}_p$ is the vector in $\mathbf{R}^p$ with all unity
components. We let
$\|z\|_P\triangleq\sqrt{z'Pz}$. $L_2[0,\infty)$ denotes the Lebesgue space of
$\mathbf{R}^k$-valued vector-functions $z(\cdot)$, defined on 
$[0,\infty)$, with the norm $\|z\|_2\triangleq
\left(\int_0^\infty\|z(t)\|^2dt\right)^{1/2}$. 

\section{Problem formulation}\label{Formulation}

\subsection{Motivating problem: synchronization of a network of
  single-input uncertain bilinear
  systems}\label{bilinear} 

Consider a collection of $N+1$ bilinear control systems-agents, labeled $0,
\ldots, N$, each described by the equation
\begin{eqnarray}
  \label{nonlin.obs.b}
   \dot{x_i}=(A_0+\rho(t)\Delta) x_i + u_i(t) +  B_{2i}w_i(t), \\
      x_i(0)=x_{i0}, \nonumber 
 \end{eqnarray}
where $x_i\in\mathbf{R}^n$ is the state of
agent $i$, $u_i\in\mathbf{R}^n$ denotes the local control input to agent
$i$ through which it can be interconnected with other agents in the
network, and $w_i\in \mathbf{R}^{r_i}$ is the
disturbance. $A_0$, $\Delta$ are constant matrices. Also, 
$\rho:[0,\infty)\to \Gamma\triangleq
[\rho^{\mathrm{min}},\rho^{\mathrm{max}}]\subset \mathbf{R}$ is a
continuous differentiable signal known to all agents.

The signal $\rho(t)$ can be thought of as control steering the entire
network to a desired behavior. 
One way to achieve this is to design such
a control signal $\rho$ for one of the agents, and then
control all other agents to track the leader; cf.~\cite{GYSS-2012}. Without
loss of generality, suppose agent 0 is 
selected to be the leader. To distinguish this agent from other agents $i$,
$i=1,\ldots,N$, its state is denoted $x$, and its evolution is
described as
 \begin{eqnarray}
  \label{plant.b}
  \dot x=(A_0+\rho(t)\Delta)x+B_{20}w_0(t), \quad x(0)=x_0.
\end{eqnarray}
The system (\ref{plant.b}) is bilinear and is governed by
the common control signal $\rho(t)$. However
for the leader we let $u_0=0$, to ensure its dynamics are not affected by the
network. 

\begin{remark}
According to one of the original viewpoints
on the gain scheduled design motivated by the objective of governing the
system to an \emph{a priori} known trajectory~\cite{Shamma-Athans-TAC},
$\rho(t)$ is assumed to be known to all agents. It can be either given
(e.g., the desired heading for the 
formation)~\cite{Shamma-Athans-TAC}, or can be computed by each agent in
the network, e.g., using nearest neighbours rules~\cite{Savkin-2004}.
\end{remark}

Suppose  each agent (\ref{nonlin.obs.b}) receives broadcast
signals $y_i$, $v_{ij}$ from the reference plant and its neighbours $j$ (cf.~\cite{U7}): 
\begin{eqnarray}\label{y.nonlin}
y_i=C_{2i}x+D_{2i}w_i, \quad
v_{ij}= H_{ij}x_j+G_{ij}w_{ij},
\end{eqnarray}
here $w_{ij}$ are
disturbances affecting the communication between $j$ and $i$. 
We wish to design a protocol for interconnecting agents over a
network, to ensure all of them track the
leader. This problem is a special
case of a more general synchronization problem introduced in the next section.
   
\subsection{Synchronization problem for a network of parameter-varying
  agents}\label{general}   

Consider a fixed directed graph 
$\mathbf{G}
= (\mathbf{V},\mathbf{E})$; $\mathbf{V}$, $\mathbf{E}$ are
the set of vertices and the set of edges (i.e, the subset of the set
$\mathbf{V}\times \mathbf{V}$), respectively. 
Without loss of generality, we let $\mathbf{V}=\{1,2,\ldots,N\}$.
The notation $(j,i)$ will denote the edge
of the graph originating at node $j$ and ending at node $i$. It is assumed that
the graph $\mathbf{G}$ has no self-loops, 
$(i,i)\not\in \mathbf{E}$. 
   
For each $i\in \mathbf{V}$, let $\mathbf{V}_i=\{j:(j,i)\in \mathbf{E}\}$
be the set of nodes supplying information to node $i$, termed as the
neighbourhood of node $i$. The cardinality of $\mathbf{V}_i$ is the
in-degree of node $i$ and is denoted $p_i$; i.e., 
$p_i$ is equal to the number of incoming edges for node $i$. Also,  $q_i$
will denote the number of outgoing 
edges for node $i$, known as the out-degree of node $i$.
According to Proposition~1 in~\cite{U6}, the attention is restricted to weakly
connected graphs.

Consider a multi-agent system, consisting of a parameter varying
nonlinear reference system  
 \begin{equation}
  \label{plant}
  \dot x=A(\rho(t))x+B_1\phi(x)+B_{20}w_0(t), \quad x(0)=x_0,
\end{equation}
and $N$ parameter varying nonlinear dynamical agents,
\begin{eqnarray}
  \label{nonlin.obs}
   \dot{x_i}=A(\rho(t)) x_i + B_1\phi(x_i) + u_i(t) +  B_{2i}w_i(t), \\
      x_i(0)=x_{i0}. \nonumber 
 \end{eqnarray}  
Here the variables $x,x_i,u_i,w_i$ have the same meaning as in
Section~\ref{bilinear}, and $\rho(t)$
is the time-varying parameter available to all agents, described
previously. The matrix-valued function $A(\cdot)$
is assumed to be continuous on the interval $\Gamma$. Also, the function
$\phi(x):\mathbf{R}^n\to \mathbf{R}^l$ satisfies the global Lipschitz condition 
\begin{equation}
  \label{Lip}
  \|\phi(x_1)-\phi(x_2)\|^2 \le (x_1-x_2)'R (x_1-x_2), \quad \forall
  x_1,x_2\in \mathbf{R}^n;
\end{equation}
where $R=R'\ge 0$.
The system (\ref{plant}) and the agents model (\ref{nonlin.obs}) include
the bilinear systems (\ref{plant.b}) and (\ref{nonlin.obs.b}) considered in
Section~\ref{bilinear} as special
cases, where $A(\rho)=A_0+\rho\Delta$ and $\phi(x)\equiv
0$. Also, we let $R=0$ in this special case. 

As stated in Section~\ref{bilinear}, we assume that direct measurements of
the reference plant (\ref{plant}) are not available, and each agent
(\ref{nonlin.obs}) must rely on broadcast
signals from the reference plant and its neighbours defined in
(\ref{y.nonlin}). It is assumed that $w_0$,
$w_i(\cdot),w_{ij}(\cdot)\in L_2[0,\infty)$, $i,j=1,\ldots,N$. 
Also,
we assume that $E_{2i}=D_{2i}D_{2i}'>0$,
$F_{ij}=G_{ij}G_{ij}'>0$ for all $i$. 

Consider the following consensus-based protocol for interconnecting the agents 
over the graph $\mathbf{G}$:
\begin{eqnarray}
u_i(t) &=& L_i(\rho(t))(y_i-C_{2i} x_i) \nonumber \\
 && +\sum_{j\in \mathbf{V}_i}K_{ij}(\rho(t))(v_{ij} - H_{ij}x_i), 
\label{u}
\end{eqnarray}
where 
$L_i(\cdot)$, $K_{ij}(\cdot)$ are matrix-valued gain functions to be
determined. 
The synchronization problem in this paper is to determine the
interconnection feedback gains $K_{ij}(\cdot)$ and 
observer gains $L_i(\cdot)$ such that $x_i(t)\to x(t)$ in  the $L_2$
sense and asymptotically. This relates our approach to the distributed
version of the observer-based synchronization problem;
cf.~\cite{NM-1997}. As a distinctive feature of our approach, we aim to
achieve synchronization while forcing 
the nodes to reach a guaranteed suboptimal level of relative $H_\infty$
disagreement, as stated in Definitions~\ref{Def1} and~\ref{Def1'} given below.
It is also worth noting that the agents in
(\ref{nonlin.obs}) have nonidentical measurement models which employ
nonidentical matrices $C_{2i}$, $D_{2i}$. 
This is an important distinction between our model and those frequently
used in the literature where all agents employ identical measurement
models; e.g., see~\cite{LDCH-2010,TTM-2013}.  

As a measure of consensus between the agents consider the disagreement 
function~(cf.~\cite{OM-2004})   
\begin{eqnarray}\label{disagr}
\Psi_\mathbf{G}(\mathbf{x})&=&\frac{1}{N} \sum_{i=1}^N \sum_{j\in
  \mathbf{V}_i}\|x_j-x_i\|^2, \quad \mathbf{x}=[x_1'~\ldots~x_N']'.
\end{eqnarray}
In this paper, two synchronization problems are considered that utilize
$\Psi_{\mathbf{G}}$ as the running cost of synchronization.
In the first problem we are concerned with achieving synchronization
with a guaranteed $H_\infty$ level of disagreement between the agents. In
the second problem, a stronger version of disagreement performance is
considered. It involves a penalty on the synchronization transient performance,
additional to the penalty on the consensus transient performance. 

Let $\mathbf{x}_0=[x_{10}',\ldots,x_{N0}']'$,  and 
\begin{eqnarray*}
\|(\mathbf{x}_0,w_0,\mathbf{w},\bar{\mathbf{w}})\|^2&\triangleq& 
\|\mathbf{1}_n\otimes x_0-\mathbf{x}_0 \|^2_P + \|w_0\|_2^2\\
&&+\frac{1}{N} \sum_{i=1}^N\left(\|w_i\|_2^2
    +\sum_{j=1}^N\|w_{ij}(\cdot)\|^2_2\right),
\end{eqnarray*}
where $P=P'>0$ is a fixed
matrix to be determined.

\begin{definition}\label{Def1}
The problem of weak robust synchronization is to determine continuous feedback
control and interconnection gain schedules $L_i(\rho)$, $K_i(\rho)$,
$\rho\in\Gamma$,  for the protocol (\ref{u}) to satisfy the following
properties:  

  \begin{enumerate}[(i)]
  \item
In the absence of uncertain perturbations, the interconnection of unperturbed systems
describing evolution of the synchronization error dynamics of each agent
$e_i\triangleq x-x_i$
must be exponentially stable. That is, 
\begin{eqnarray*}
\|e_i(t)\|^2\triangleq \|x_i(t)-x(t)\|^2\le c e^{-\omega t}, \quad (\exists
c,\omega>0). 
\end{eqnarray*}

\item
In the presence of uncertain perturbations, the protocol (\ref{u}) must ensure 
a certain level of $H_\infty$ consensus performance in
the following sense 
\begin{eqnarray}\label{objective.i.1}
&&\sup_{\|(\mathbf{x}_0,w_0,\mathbf{w},\bar{\mathbf{w}})\|\neq 0}
\frac{\int_0^\infty \Psi_\mathbf{G}(\mathbf{e}(t))dt}
{\|(\mathbf{x}_0,w_0,\mathbf{w},\bar{\mathbf{w}})\|^2} \le \gamma^2.
\end{eqnarray}
Here,  $\gamma>0$ is a given constant, and $\mathbf{e}=[e_1',\ldots,e_N']'$.  

\item
All agents synchronize asymptotically, i.e., for all $w_0,w_i,w_{ij}\in L_2[0,\infty)$, 
\begin{eqnarray}
&&\lim_{t\to\infty}\sum_{i=1}^N\|x(t)-x_i(t)\|^2=0 .
\label{convergence} 
\end{eqnarray}
  \end{enumerate}
\end{definition}

\begin{definition}\label{Def1'}
Let $Q_1,\ldots,Q_N$
be given symmetric positive definite matrices, and  $Q\triangleq
\mathrm{diag}[Q_1,\ldots,Q_N]$.   
The problem of strong robust synchronization is to
finding  continuous feedback control and interconnection gain 
schedules $L_i(\rho)$, $K_i(\rho)$, $\rho\in\Gamma$,  for the protocol
(\ref{u}) to satisfy conditions (i), (iii) of Definition~\ref{Def1},
and the following condition, which replaces (\ref{objective.i.1}) in (ii):
\begin{eqnarray}\label{objective.i.2}
\sup
\frac{\int_0^\infty \left(\frac{1}{N}\mathbf{e}(t)'Q\mathbf{e}(t)
+\Psi_\mathbf{G}(\mathbf{e}(t))\right)dt}
{\|(\mathbf{x}_0,w_0,\mathbf{w},\bar{\mathbf{w}})\|^2} 
\le \gamma^2.
\end{eqnarray}
where the $\sup$ is over the set $\{\|(\mathbf{x}_0,w_0,\mathbf{w},\bar{\mathbf{w}})\|\neq 0\}$.
\end{definition} 

Condition (\ref{objective.i.2}) guarantees  
$H_\infty$ filtering performance $\int_0^\infty\mathbf{e}(t)'Q\mathbf{e}(t)dt
\le\gamma^2 N\|(\mathbf{x}_0,w_0,\mathbf{w},\bar{\mathbf{w}})\|^2$, which is
not guaranteed by conditions in Definition~\ref{Def1}. Hence the term
strong synchronization. Also, the novel aspect in both definitions
concerns the asymptotic synchronization property in condition (iii);
an $L_2$ convergence was claimed in~\cite{U6}.      

\section{The main results}\label{main.results}

The derivation of the main result of the paper will proceed in several
steps, following the general scheme of stability and robustness preserving
interpolated
gain-scheduling~\cite{Stilwell-Rugh-99,Stoustrup-Komareji,YUMP1}. First, we 
revisit the results in \cite{U6} for a fixed parameter case and 
a more general class of agents under consideration in
this paper. Recall that 
in \cite{U6}, an ideal communication between the agents was assumed, whereas
in this paper we allow for a more general situation where the messages
between the agents are subject to disturbance. Next,
based on this extension, a synchronization result will be established 
for a class of parameter-varying systems with a small-scale parameter
variation. While technically simple,
this extension will serve as the basis for the derivation of interpolated
feedback schedules for a more general class of parameter-varying systems
with bounded rate of parameter variations, which is the main result of this
paper. 

\subsection{Synchronization of fixed parameter systems} 
Let us fix $\rho\in \Gamma$ and consider the fixed parameter version of the
uncertain reference system (\ref{plant}),
 \begin{equation}
  \label{plant.fixed}
  \dot x=A(\rho)x+B_1\phi(x) +B_{20}w_0+\psi(t,x), \quad x(0)=x_0,
\end{equation}
and the corresponding $N$ uncertain fixed-parameter dynamical agents 
\begin{eqnarray}
   \dot{x_i}&=&A(\rho) x_i + B_1\phi(x_i) + \psi(t,x_i) \nonumber 
   \\ && + u_i(t) + B_{2i} w_i(t),  \qquad  x_i(0)=x_{i0}.
  \label{nonlin.obs.fixed}
 \end{eqnarray}
Compared with  (\ref{nonlin.obs}), we have introduced
an additional uncertainty term to the reference plant and the equations of
agents' dynamics. The motivation for this will become clear later, when we 
will consider small parameter variations in the agents and the reference
plant. Such small variations can be treated as an additional
norm-bounded uncertainty, capturing the mismatch between fixed system
parameters used in the protocol design, and the true system parameters. It
will be shown that the  size of this mismatch can be characterized in terms
of a uniform norm bound condition, such as   
\begin{eqnarray}
\|\psi(t,x)-\psi(t,x_i)\|^2&\le& \alpha^2 \|e_i\|^2,  \label{wv.constr} 
\end{eqnarray}
where $\alpha>0$ is a constant. 
 
We now present a sufficient condition for the existence of a fixed
parameter version of the protocol (\ref{u}) which ensures that the fixed
parameter 
systems (\ref{nonlin.obs.fixed}) achieve strong synchronization. 
Given constants $\delta_i>0$ and matrices $Q_i=Q_i'>0$, $i=1,\ldots, N$,
introduce the following coupled 
Linear Matrix Inequalities (LMIs) in scalar variables $\tau_i>0$,
$\theta_i>0$ and matrix variables $X_i=X_i'>0$: 
\begin{eqnarray}
\left[
\begin{array}{ccc}
S_i(\rho)+\tau_i R+\theta_i\alpha^2I & \star & \star  \\
T_i' & -\Upsilon_i & * \\
X_i  & 0 & -\theta_iI
\end{array}\right]<0, \label{T4.LMI.1} 
\end{eqnarray}
where $T_i=\left[\begin{array}{cccc}
X_iB_{20} & X_iB_2(I-D_{2i}'E_i^{-1}D_{2i}) & X_iB_1 & \Xi_i
\end{array}\right],$
\begin{eqnarray}
&&
S_i(\rho)\triangleq X_i(A(\rho)+\delta_iI+B_{2i}D_{2i}'E_{2i}^{-1}C_{2i})
\nonumber \\
&&\phantom{S_i(\rho)}
+(A(\rho)+\delta_iI+B_{2i}D_{2i}'E_{2i}^{-1}C_{2i})'X_i+(p_i+q_i)I
\nonumber \\
&& \phantom{S_i(\rho)}
-\gamma^2C_{2i}'E_{2i}^{-1}C_{2i}- \gamma^2\sum_{j\in
  \mathbf{V}_i}H_{ij}'F_{ij}^{-1}H_{ij}+Q_i, \nonumber \\
&&
\Xi_i=\left[\begin{array}{ccc}\gamma^2H_{ij_1}'F_{ij_1}^{-1}H_{ij_1}\!-I & \ldots & \gamma^2H_{ij_{p_i}}'F_{ij_{p_i}}^{-1}H_{ij_{p_i}}\!-I
  \end{array}
  \right], \nonumber \\
&& \Upsilon_i=\mathrm{diag}\left[\gamma^2I,~\gamma^2I,~\tau_i I,~Z_i\right],
\nonumber \\ 
&&
Z_i=\mathrm{diag}\left[\frac{2\delta_{j_1}}{q_{j_1}+1} X_{j_1},~\ldots,
  ~ \frac{2\delta_{j_{p_i}}}{q_{j_{p_i}}+1} X_{j_{p_i}} \right],
\label{SiXiZi}
\end{eqnarray}
$j_1,\ldots, j_{p_i}$ are the elements of the neighbourhood set
$\mathbf{V}_i$. 

\begin{lemma}\label{T.aux}
Let $\rho\in\Gamma$ be given and fixed. 
Suppose the graph $\mathbf{G}$, the matrices $Q_i=Q_i'>0$ and the constants
$\gamma>0$ and 
$\delta_i>0$ are such that the coupled LMIs (\ref{T4.LMI.1}) corresponding
to the given $\rho$ 
are feasible. Consider a collection of feasible triples
$(\tau_{i,\rho},\theta_{i,\rho}, X_{i,\rho})$, $i=1,\ldots,N$,
 and define
\begin{eqnarray}
 K_{ij}(\rho)&=&\gamma^2X_{i,\rho}^{-1}H_{ij}'F_{ij}^{-1}, \label{Kim}
 \\ 
 L_i(\rho)&=&(\gamma^2 X_{i,\rho}^{-1}C_{2i}'-B_{2i}D_{2i}')E_{2i}^{-1}.      \label{Lim}
\end{eqnarray}
The network of fixed parameter agents (\ref{nonlin.obs.fixed}), 
 equipped with the
protocols (\ref{u}) with the coefficients defined in (\ref{Kim}),
(\ref{Lim}) solves the fixed parameter version of the strong robust
synchronization problem in Definition~\ref{Def1'}, with $\rho(t)\equiv
\rho$. The matrix $P$ in condition (\ref{objective.i.2}) corresponding to
this solution is $P=\mathrm{diag}[\frac{1}{N}X_{i,\rho}]$.   
\end{lemma}

The proof of this lemma, as well as the proofs of other results in this
paper is given in the Appendix. The main idea of the proof is to show that the
interconnected system describing dynamics of the synchronization errors 
associated with the reference (\ref{plant.fixed}) and the multi-agent
system (\ref{nonlin.obs.fixed}) has the properties of vector
dissipativity~\cite{HCN-2004} with respect to the vector storage function
$[V_1(e_1)~\ldots~V_N(e_N)]'$, $V_i(e_i)=e_i'X_{i,\rho}e_i$, and a suitably
defined vector of supply rates.  

\begin{remark}\label{Rem.coupled.LMI}
The LMIs in (\ref{T4.LMI.1}) are coupled for neighbouring
agents. In~\cite{U6} we noted that in principle LMIs of this type can be
solved in a distributed manner. Due to space limitation, we refer the
reader to~\cite{U6} for details.
\end{remark}

\subsection{Synchronization under small parameter variations}

The protocol (\ref{u}) defined in the previous section for the fixed
parameter network 
can be used in the derivation of a synchronization protocol for the
parameter varying multi-agent system (\ref{plant}), (\ref{nonlin.obs}), provided parameter
variations are sufficiently small, i.e. $\rho(t)\approx \rho^0$ $\forall
t\ge 0$. In this case small variations of the matrix
$A(\cdot)$ can be treated as parameter mismatch disturbances, and the
parameter varying system (\ref{nonlin.obs}) and the reference (\ref{plant})
can be regarded as a perturbation of a corresponding fixed-parameter
system. To formalize this observation, let us fix $\rho^0 \in \Gamma$, and
define  
\[
\psi(t,x)=(A(\rho(t))-A(\rho^0))x.
\]
The robust fixed parameter synchronization
protocol of the form (\ref{u}) can now be designed based on the representation
(\ref{plant.fixed}),
(\ref{nonlin.obs.fixed}) using Lemma~\ref{T.aux}. 
This leads to the following result
about
synchronization of the system (\ref{plant}), (\ref{nonlin.obs}) under small
parameter variations.

\begin{theorem}\label{T.aux.small}
Let $\rho^0\in\Gamma$ be fixed. Suppose that for all $t>0$
\begin{equation}
  \label{Anorm}
  (A(\rho(t))-A(\rho^0))'(A(\rho(t))-A(\rho^0))\le \alpha^2 I.
\end{equation}
Suppose the graph $\mathbf{G}$ and the constants $\gamma>0$ and
$\delta_i>0$ are such that the coupled LMIs (\ref{T4.LMI.1}) with
$\rho=\rho^0$ are feasible, and let
$(\tau_{i,\rho^0},\theta_{i,\rho^0}, X_{i,\rho^0})$, be a corresponding
collection of feasible triples, $i=1,\ldots,N$. Then the network of agents
(\ref{nonlin.obs}) equipped with the 
protocols (\ref{u}),  (\ref{Kim}), (\ref{Lim}), where $\rho=\rho^0$, 
solves the problem of strong robust synchronization in
Definition~\ref{Def1'}. The matrix $P$ in conditions (\ref{objective.i.1})
and ~(\ref{convergence}) corresponding to this solution is
$P=\mathrm{diag}[\frac{1}{N}X_{i,\rho^0}]$.
\end{theorem}
 
\subsection{Disagreement gain preserving interpolation of synchronization
  protocols} 

Naturally, it may be difficult to cover the entire
interval $\Gamma$ using a single condition (\ref{Anorm}) while ensuring
that the coupled LMIs (\ref{T4.LMI.1}) are feasible for the selected
$\alpha$. The idea behind the gain-scheduling approach in this paper is to
cover the interval $\Gamma$ using smaller intervals for which
conditions (\ref{Anorm}) and (\ref{T4.LMI.1}) hold simultaneously.  
First a collection of constants $\alpha_k>0$ and `grid points'
$\Gamma_0=\{\rho^k,k=1,\ldots,M\}$ is selected so that for any
$\rho\in\Gamma$ there exists at least one point $\rho^k$ with the property
\begin{equation}
  \label{Anorm.1}
  (A(\rho)-A(\rho^k))'(A(\rho)-A(\rho^k))\le \alpha_k^2 I.
\end{equation}
Let $U_k$ be the largest connected neighbourhood of $\rho^k$ consisting of
all $\rho\in\Gamma$ for which (\ref{Anorm.1}) holds. The grid points
$\rho^k$ and the constants $\alpha_k$ must be selected so 
that $\Gamma\subseteq \cup_{k=1}^MU_k$. Next, using
Lemma~\ref{T.aux}, we compute the synchronization 
protocol (\ref{u}) for the uncertain parameter-varying agents 
plants (\ref{nonlin.obs.fixed}) for each fixed
$\rho^k$. The robustness properties of this protocol stated in 
Theorem~\ref{T.aux.small} guarantee synchronization for every fixed $\rho\in
\Gamma$. This property establishes an analog to the stability covering
condition in \cite{Stilwell-Rugh-99}.

Assuming that for every $\alpha=\alpha_k$ and $\rho=\rho^k$, $k=1,\ldots,M$,
the LMIs (\ref{T4.LMI.1}) are 
feasible, the above procedure guarantees that 
for each fixed $\rho\in\Gamma$, a synchronization protocol (\ref{u}) can
be assigned to the system (\ref{nonlin.obs}) by selecting one of the 
protocols corresponding to an index $k\in \{k: \rho\in U_k\}$. 
However, when applied to the parameter-varying
system (\ref{plant}) directly, this procedure will result in the
matrix-valued functions $L_i(\rho(t))$ and $K_{ij}(\rho(t))$ having jumps
at the time instants when the trajectory of the parameter $\rho(t)$ exits
the set $U_k$ and enters the set $U_{k+1}$. As a result, the control
signals $u_i$ will become discontinuous and will generate transients that
usually have an adverse effect on the system performance. To overcome these
effects, we propose a 
{\em  continuous} interpolation of the 
fixed-parameter synchronization protocols, following the interpolation
approach in 
\cite{Stilwell-Rugh-99,YUMP1,Stoustrup-Komareji}. The aim of our interpolation
technique is to preserve the property of
interpolants to guarantee a desired level of the relative $H_\infty$
disagreement between the agents. 

To explain the idea of the proposed interpolation, let us consider an
arbitrary fixed $\rho\in\Gamma$, and the collection of constants
$\alpha_k>0$, $k =1,\ldots,M$, and grid 
points $\Gamma_0$ discussed above. Since $\rho\in\Gamma$ is fixed, there
must exist $k$ such that $\rho\in U_k$. 
Let $(\tau_{i,\rho^k},\theta_{i,\rho^k}, X_{i,\rho^k})$, $i=1,\ldots,N$,
be a feasible triple of the LMIs
(\ref{T4.LMI.1}) corresponding to this $k$. The following lemma follows
from (\ref{T4.LMI.1}) and  (\ref{Anorm.1}) using the Schur complement.

\begin{lemma}\label{reduced.LMI}
Suppose (\ref{Anorm.1}) holds. The collection of matrices and constants
$\{\tau_{i,\rho^k}, X_{i,\rho^k},i=1,\ldots,N\}$, 
composed using a feasible solution of the LMI (\ref{T4.LMI.1})
corresponding to $\alpha=\alpha_k$, also satisfies the following coupled
LMIs in $X_i=X_i'>0$, $\tau_i>0$, $i=1,\ldots,N$:  
 \begin{eqnarray}
\left[
\begin{array}{cc}
S_i(\rho)+\tau_i R & \star \\
T_i' & -\Upsilon_i 
\end{array}
\right]<0,
 \label{T4.LMI.k} 
\end{eqnarray}
where $S_i(\rho)$, $T_i$ and $\Upsilon_i$ are the same as in
(\ref{T4.LMI.1}).
\end{lemma}

We now define interpolated
gains for the protocols (\ref{u}), as follows. Suppose the collection of
positive constants $\alpha_k$ and the grid points $\Gamma_0$ has the
following properties:  
\begin{eqnarray}
  \label{Anorm.2.k}
&&(A(\rho)-A(\rho^k))'(A(\rho)-A(\rho^k))\le \alpha_k^2 I, \\
&&\hspace{4cm}\mbox{if } \rho^k\le \rho<\bar \rho^k, \nonumber  \\ 
  \label{Anorm.2.k+1}
&&(A(\rho)-A(\rho^{k+1}))'(A(\rho)-A(\rho^{k+1}))\le \alpha_{k+1}^2 I,
 \\
&&\hspace{4cm}\mbox{if } \underline{\rho}^{k+1}< \rho\le \rho^{k+1}, \nonumber
\end{eqnarray}
where $\rho^k < \underline{\rho}^{k+1} < \bar \rho^k < \rho^{k+1}$. 
In particular, this implies that $[\underline{\rho}^{k+1},\bar \rho^k]\subset
U_k\cap U_{k+1}$, and both (\ref{Anorm.2.k}) and (\ref{Anorm.2.k+1}) hold
for $\rho\in [\underline{\rho}^{k+1},\bar \rho^k]$. 

For every $\rho\in \Gamma$, select $k,k+1$ such that
$\rho\in[\rho^k,\rho^{k+1}]$, and define $\lambda=\frac{\bar\rho^k-\rho}{\bar\rho^k-\underline\rho^{k+1}}$,
\begin{eqnarray}
  \label{X.interp}
X_{i,\rho}&=&\begin{cases}
X_{i,\rho^k},       & \hspace{-2cm}\rho\in [\rho^k,\underline{\rho}^{k+1}], \\ 
X_{i,\lambda}\triangleq \lambda X_{i,\rho^k}+(1-\lambda)X_{i,\rho^{k+1}}, \\
& \hspace{-2cm}\rho\in [\underline{\rho}^{k+1},\bar \rho^k], \\
X_{i,\rho^{k+1}},     & \hspace{-2cm}\rho\in [\bar\rho^k,\rho^{k+1}],
\end{cases} \\
  \label{tau.interp}
\tau_{i,\rho}&=&\begin{cases}
\tau_{i,\rho^k},       & \hspace{-2cm}\rho\in [\rho^k,\underline{\rho}^{k+1}], \\ 
\tau_{i,\lambda}\triangleq \lambda\tau_{i,\rho^k}+(1-\lambda)\tau_{i,\rho^{k+1}}, \\
& \hspace{-2cm}\rho\in [\underline{\rho}^{k+1},\bar \rho^k], \\
\tau_{i,\rho^{k+1}},     & \hspace{-2cm}\rho\in [\bar\rho^k,\rho^{k+1}],
\end{cases} \\
  \label{K.interp}
 K_{ij}(\rho)&=&\gamma^2X_{i,\rho}^{-1}H_{ij}'F_{ij}^{-1} \\
L_i(\rho)&=&(\gamma^2X_{i,\rho}^{-1}C_{2i}'-B_{2i}D_{2i}')E_{2i}^{-1}.
\label{L.interp}
\end{eqnarray}
Note that the matrix gains $K_{ij}$, $L_i$ defined above, are continuous
and well defined functions on $\Gamma$, since $X_{i,\lambda}>0$ for all
$\lambda \in[0,1]$, and hence $X_{i,\lambda}^{-1}$ is well defined. 

The following theorem is the main result of the paper. 
Let $\bar\Gamma$ be the set consisting of all the
corner points $\underline\rho^{k+1}$, $\bar\rho^k$, which lie inside
$\Gamma$. Without loss of generality, we assume that $\rho(0)\not\in
\bar\Gamma$, and that the set $\{t\colon \rho(t)\in\bar\Gamma\}$ has zero
Lebesgue measure.  

\begin{theorem}\label{main.T}
  Suppose
\begin{equation}
  \label{rate.condition}
  \sup_{t\ge 0}|\dot\rho|\le 
  \min_{i}\left\{\lambda_{\min}(Q_i)\left[\sup_{k}\frac{\left\|X_{i,\rho^{k+1}}-X_{i,\rho^{k}}\right\|}{\bar \rho^k-\underline\rho^{k+1}} \right]^{-1}\right\}.
\end{equation}
Then the network of agents (\ref{nonlin.obs}) equipped with the 
protocols (\ref{u}),  (\ref{K.interp}), (\ref{L.interp}),  
solves the weak robust synchronization problem in Definition~\ref{Def1}. The
matrix $P$ in condition (\ref{objective.i.1})
corresponding to this solution is
$P=\mathrm{diag}[\frac{1}{N}X_{i,\rho(0)}]$.
\end{theorem}

If the inequality in (\ref{rate.condition}) is replaced with a strict
inequality, then strong synchronization can be established.

\begin{corollary}\label{Cor}
  Suppose there exists $\eta\in [0,1)$ such that 
\begin{equation}
  \label{rate.condition.strong}
  \sup_{t\ge 0}|\dot\rho|\le \eta
  \min_{i}\left\{\lambda_{\min}(Q_i)\left[\sup_{k}\frac{\left\|X_{i,\rho^{k+1}}-X_{i,\rho^{k}}\right\|}{\bar \rho^k-\underline\rho^{k+1}} \right]^{-1}\right\}.
\end{equation}
Then the network of agents (\ref{nonlin.obs}) equipped with the 
protocols (\ref{u}),  (\ref{K.interp}), (\ref{L.interp}),  
solves the strong robust synchronization problem in Definition~\ref{Def1'}, with
$Q$ in (\ref{objective.i.2}) replaced with $(1-\eta)Q$. The matrix $P$ is
defined in the same way as in Theorem~\ref{main.T}. 
\end{corollary}

\begin{remark}\label{rem.gamma}
It is worth noting that the variable $\gamma^2$ enters linearly
in~(\ref{T4.LMI.1}). This allows for optimization over $\gamma^2$. Such
optimization can be carried out for each grid point $\rho^k$, and then the
largest value out of obtained $\gamma^2_k$ should be selected. Maximization
over $k$ is justified
because  
if for a fixed $\rho=\rho_k$ the LMIs (\ref{T4.LMI.1}) admit a solution for
$\gamma=\gamma_k$, then the same solution is feasible for these LMIs with any
$\gamma>\gamma_k$.  
\end{remark}

\section{Application to synchronization of single-input bilinear
  systems}\label{sec.bi} 

We now apply the results of the previous section to the problem of
synchronization of bilinear systems described in Section~\ref{bilinear},
where the matrix $A(\rho)$ is  linear, 
$A(\rho)=A_0+\rho\Delta$, $A_0$, $\Delta$ are constant matrices.
Owing to the affine structure of $A(\rho)$, the construction
of the synchronization protocol can be further simplified, as presented below.

Let $\alpha>0$ and 
grid points $\Gamma_0=\{\rho^k,k=1,\ldots,M\}$ be selected so that
$ 
|\rho^k-\rho^{k+1}|<\frac{\alpha}{\sigma(\Delta)},  
$ 
$\sigma(\Delta)$ is the largest singular value of $\Delta$, and
$\rho^1\le \rho^{\mathrm{min}}$, $\rho^M\ge \rho^{\mathrm{max}}$. Then the
condition (\ref{Anorm.1}) trivially holds for any
$\rho\in U_k=[\rho^{k-1},\rho^{k+1}]$, with $U_1=[\rho^1,\rho^2]$,
$U_M=[\rho^{M-1},\rho^M]$.  Also, $\Gamma\subseteq \cup_{k=1}^MU_k$. 

Suppose $\alpha$ is chosen so that for every $\alpha$ and $\rho=\rho^k$,
$k=1,\ldots,M$, the LMIs  (\ref{T4.LMI.1}) in the variables $(\theta_i,X_i)$
$X_i=X_i'>0$, with $R_i=0$, 
are feasible. We allow for $R_i=0$ in this section since here we consider the
special case $\phi(\cdot)\equiv 0$. 
The synchronization protocol (\ref{u}) can then computed for each
$\rho^k$ using Lemma~\ref{T.aux}, in which we let $R_i=0$.
Furthermore, it is clear that for all $\rho\in\Gamma$, there exist two
adjacent grid points $\rho^k,\rho^{k+1}$ such that $\rho\in U_{k}\cap
U_{k+1}$. Specifically, these $\rho^k,\rho^{k+1}$ are the endpoints of the
interval $[\rho^k,\rho^{k+1}]$ to which $\rho$ belongs.

These considerations allow us to simplify the procedure of the previous
section to construct interpolated gains for the protocol
(\ref{u}). For a $\rho$ such that $\rho\in [\rho^k,\rho^{k+1}]$, let 
\begin{eqnarray}
  \label{X.interp.b}
X_{i,\rho}&=&\frac{\rho^{k+1}-\rho}{\rho^{k+1}-\rho^k} X_{i,k}+
\frac{\rho-\rho^k}{\rho^{k+1}-\rho^k} X_{i,k+1}.
\end{eqnarray}

The following theorem specializes Theorem~\ref{main.T} and
Corollary~\ref{Cor} to the problem of synchronization of bilinear systems. 

\begin{theorem}\label{main.T.b}
  Suppose condition (\ref{rate.condition}), where $X_{i,k}$ are defined from
  (\ref{T4.LMI.1}) with $R_i=0$, holds. Then the network of agents (\ref{nonlin.obs.b})
  equipped with the protocols (\ref{u}),  (\ref{K.interp}),
  (\ref{L.interp}), solves the weak robust synchronization problem in
  Definition~\ref{Def1}.  Furthermore, if a stronger condition 
(\ref{rate.condition.strong}) holds, then
the network of agents (\ref{nonlin.obs.b}) equipped with the 
protocols (\ref{u}),  (\ref{K.interp}), (\ref{L.interp}),  
solves the strong robust synchronization problem in Definition~\ref{Def1'}, with
$Q$ replaced with $(1-\eta)Q$. In both cases, the matrix $P$ in these
definitions is obtained to be $P=\mathrm{diag}[\frac{1}{N}X_{i,\rho^0}]$.
\end{theorem}

\section{Example: Master-slave synchronization of chaotic
  oscillators}\label{example} 
To illustrate the results of the paper, consider the master-slave
synchronization problem for a set of five 2nd-order bilinear systems with a
so-called unified chaotic system~\cite{LCCC-2002} as a reference,  
\begin{eqnarray}
  \dot x^{(1)}&=& (25\vartheta+10) (x^{(2)}- x^{(1)}), \nonumber \\
  \dot x^{(2)}&=& (28-35\vartheta)x^{(1)}+(29\vartheta-1)x^{(2)}-\rho(t) x^{(1)},  \nonumber \\
  \dot \rho   &=&
  -\frac{8+\vartheta}{3}\rho(t)+x^{(1)}x^{(2)}.      \label{lor-master} 
\end{eqnarray}
This system exhibits chaotic dynamics for all values of
$\vartheta\in[0,1]$~\cite{LCCC-2002}. Although the system is nonlinear, the
first two equations can be regarded as a single-input bilinear system, with
$\rho(t)$ interpreted as a control variable. This subsystem has the
form (\ref{plant}) in which $w_0=0$,  
$\phi(x)=(\vartheta-\frac{1}{2})[x^{(1)}~x^{(2)}]'$, and 
\[
A(\rho)=\left[\begin{array}{cc}
  -22.5 &  22.5 \\
   10.5 &  13.5
\end{array}\right]+\rho\left[\begin{array}{cc}
0 & 0\\
-1 & 0
\end{array}\right], \quad 
B_1=\left[\begin{array}{cc}
   -25 &   25\\
   -35 &   29
\end{array}\right].
\]
To complete the analogy
with (\ref{plant}), we randomly generated the matrix $B_{20}$ to be
$B_{20}=[0.0806~0.0232]'$, Clearly, (\ref{Lip}) holds with $R=0.25I$
irrespective of $\vartheta$. 

The network to be synchronized consists of controlled systems of the form
(\ref{lor-master}). However, in accordance with the master-slave approach
to synchronization, the signal $\rho$ is supplied by the master system
(\ref{lor-master}), rather then being generated within the slave system. 
As a result, each slave system can be regarded as a 2nd-order system of the form
(\ref{nonlin.obs}) governed by the signal $\rho(t)$ generated
by the reference system (\ref{lor-master}). We assume $B_{2i}=B_{20}$ for
all agents.  

It is easy to verify using simulations that if $u_i\equiv 0$, the agents
systems do not synchronize to the reference system.
This motivates us to introduce an additional control of the form (\ref{u})
to achieve synchronization. For this example, a simple ring
structure of the network was chosen, so that agent $i$ can receive
information from agent $i-1$ and can forward its 
state to agent $i+1$. That is, $\mathbf{V}_i=\{i-1\}$ for $i=2,\ldots$, and
$\mathbf{V}_1=\{N\}$. In this example, $N=5$.  

Since the framework of the paper allows for synchronization via imperfect
measurements and imperfect communication, we let $D_{2i}=0.01$,
$G_{i,i-1}=0.2$ and randomly selected a set of matrices $C_{2i}$,
$H_{i,i-1}$ to provide each agent with partial  measurements of the
reference plant and the partial information about the states of its
neighbours. Also, the reference system (\ref{lor-master}) was simulated 
on the interval $[0,100]$, with the initial conditions $[0.3~0.3~20]'$ and
several values of $\vartheta$, to determine the bound on $|\dot\rho|$. It was
found that on this time interval, $|\dot\rho|\le 523.3044$ for the tested
values of $\vartheta$.  

To determine the interval $\Gamma$, we used the theoretical
bound on trajectories of the Lorenz system $|\rho(t)-r|\le
\frac{b}{2\sqrt{b-1}}r$, where
$r=28-35\vartheta>0$~\cite{Swinnerton-Dyer-2001}.  Letting
$\vartheta=0.5$ (the nominal system), we found $\Gamma=[0,56.2726]$.     
Next, 
11 evenly spaced grid points were chosen as the set $\Gamma_0$,
with $\rho^0=0$, $\rho^{11}=56.2726$,
and the LMI
optimization problem $\min\gamma^2$ subject to the LMI (\ref{T4.LMI.1}) was
solved at each grid point, with $\delta=0.12$, $\alpha_k^2=32.3026$, and
$Q=17\times I$, to obtain the matrices $X_{i,\rho^k}$. 
These parameters were chosen to ensure that conditions
(\ref{Anorm.2.k}), 
(\ref{Anorm.2.k+1}) hold with $\bar\rho^k=\rho^{k+1}$ and
$\underline\rho^{k+1}=\rho^k$. Clearly,
$\rho(0)=20\not\in\Gamma_0$, as required in Theorem~\ref{main.T}.
Note that although the value of $\vartheta$ is assumed to be known in this
example, we chose to use the matrix $R=0.25 I$ which provides the upper bound on
the Lipschitz constant of $\phi$ over $\vartheta$. This allows us to design
the protocol (\ref{u}) to be independent of the value of $\vartheta$,
thus providing additional robustness.   
 
We verified that the conditions of Theorem~\ref{main.T} were satisfied in
this example, except for the rate bound condition
(\ref{rate.condition}). Unfortunately, this condition was difficult to
satisfy in this example. Despite this, our simulations confirmed
synchronization. This indicates that condition (\ref{rate.condition}) is
conservative. 

The obtained matrices $X_{i,\rho^k}$ were used to  schedule the
coefficients  for the protocol (\ref{u}) according to (\ref{K.interp}) and
(\ref{L.interp}). Here we implicitly used the observation that if the set
of LMIs (\ref{T4.LMI.1}) admits a solution for $\gamma=\gamma_k$, then
the same solution is feasible for these LMIs with any
$\gamma>\gamma_k$; see Remark~\ref{rem.gamma}. Hence, the upper bound on
the $H_\infty$ disagreement 
gain, guaranteed by the interpolated gain-scheduled synchronization
protocol constructed in this example will be $\max_{k}\gamma_k^2=
1.9729$. 

\begin{figure}
  \centering
% \subfigure[\label{open-loop}]{%
% \includegraphics[width=0.45\textwidth]{lorchen1.notcontrolled.eps}
% }
\subfigure[\label{(a)}]{%
\includegraphics[width=0.39\textwidth]{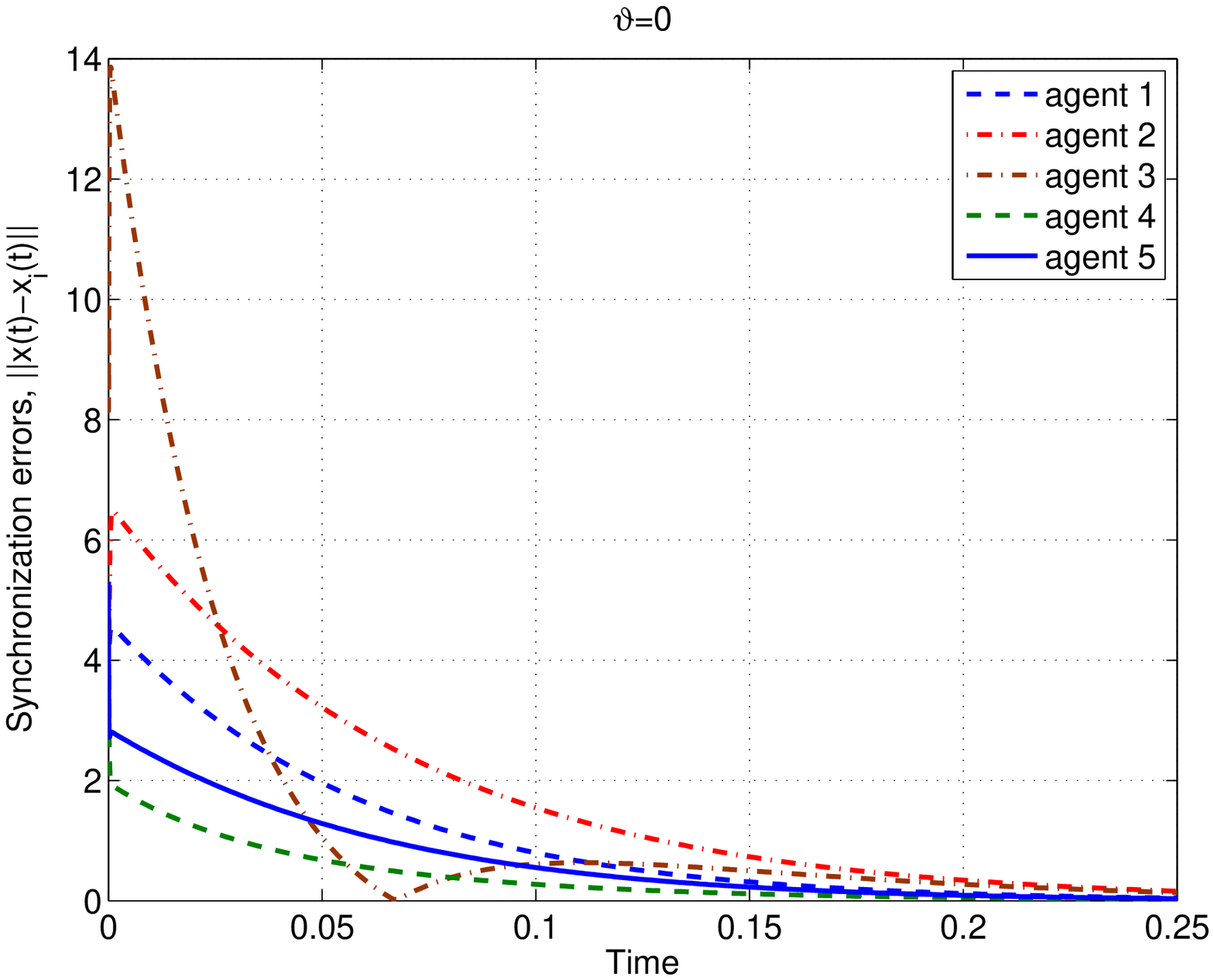}
}

\subfigure[\label{(b)}]{%
\includegraphics[width=0.39\textwidth]{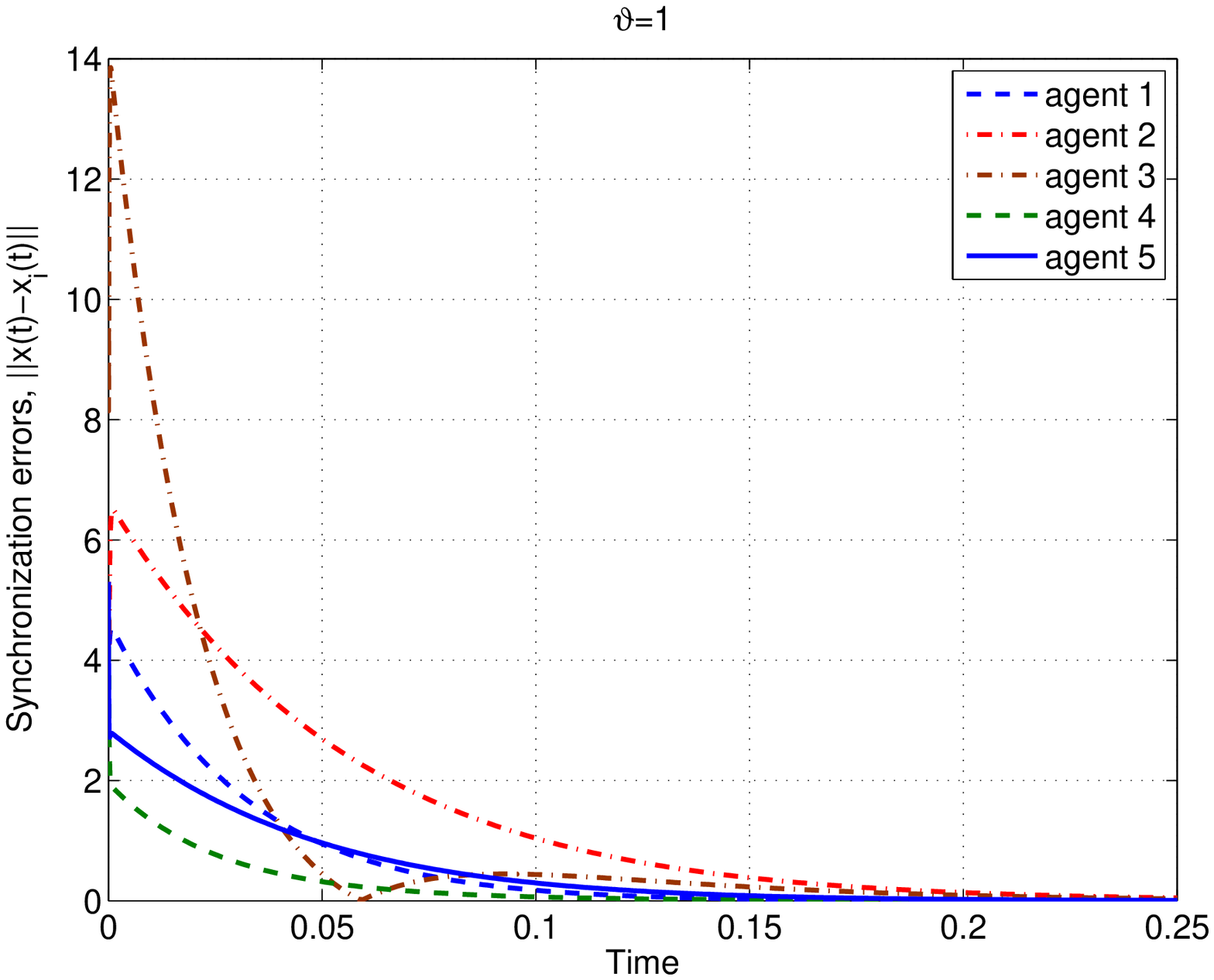}
}
\caption{Synchronization errors $\|x-x_i\|$ versus
    time: % (a) Decoupled systems ($\vartheta=0$,
      % $u_i\equiv 0$); 
    (a) Interconnected systems, $\vartheta=0$; (b) 
    Interconnected systems, $\vartheta=1$.}
\label{Synch.erros}    
\end{figure}

To verify the design, the interconnected system  was simulated on
the time interval $[0,100]$. The plots of the errors $\|x-x_i\|$ for
$\vartheta=0$ and $\vartheta=1$ are shown in Figure~\ref{Synch.erros}. It
was observed in our simulations that
all synchronization errors converged to 0, as was predicted by
Theorem~\ref{main.T}, even though the rate bound (\ref{rate.condition})
failed to satisfy in this example. It is also worth noting that despite
value of $\rho(t)$ varies substantially in this example, the error dynamics
exhibit no sign of performance degrading transients  while $\rho(t)$ varies. This reflects favourably
on the proposed interpolated synchronization protocol.

% Data from the file load data_lor_chen.mat
% Obtained using lor_chen_1.m
% and simulated using chen_sync_1.mdl
% 19/08/2013
%
     
 \section{Conclusion}
The paper has extended the
gain-scheduling via interpolation technique to the class of synchronization
problems for large-scale systems consisting of parameter varying agents
with a Lipschitz continuous nonlinearity.  
The results
have been applied to synchronization of multiagent network of bilinear
systems.

It has been observed in our previous work~\cite{U6} that the
synchronization scheme proposed in that paper tends to use high gain
observers to achieve synchronization. Large
Lipschitz constants associated with system nonlinearities 
are likely to contribute to
conservatism of the conditions in~\cite{U6}. Consequently, this could be
one reason for the algorithm~\cite{U6} to yield high observer and
interconnection gains. LPV modelling along the system trajectory may
potentially reduce the size of the nonlinearity, and hence may lead to
reduced gains required for synchronization. The results in this paper serve
as a starting point for investigation into this hypothesis.
   
\section{Appendix}

\subsection{Proof of Lemma~\ref{T.aux}}

Consider a collection of interconnected systems describing dynamics of the
synchronization errors associated with the system (\ref{plant.fixed}),
(\ref{nonlin.obs.fixed}) and the protocol (\ref{u}),
\begin{eqnarray}
   \dot{e}_i 
  &=& (A(\rho) - L_i(\rho)C_{2i})e_i+B_1\xi_i(t)+\zeta_i(t)\nonumber \\ 
  &+&\sum_{j\in\mathbf{V}_i}K_{ij}(\rho)(H_{ij}(e_j-e_i)-G_{ij}w_{ij})
             \nonumber \\  
 &+&+B_{20}w_0(t)-(B_{2i}+L_i(\rho)D_{2i}){w_i}(t),
  \label{enon.fixed.1} \qquad \\
  e_i(0)&=&x_0-x_{i0}, \nonumber\\
\xi_i(t)&=& \phi(x(t))-\phi(x_i(t)), \nonumber\\
 \zeta_i(t)&=& \psi(t,x(t))-\psi(t,x_i(t)). \label{vi.fixed} 
\end{eqnarray}

Using the constants $\tau_{i,\rho}>0$, $\theta_{i,\rho}>0$ and the matrix
$X_{i,\rho}$ obtained from the LMIs~(\ref{T4.LMI.1}), define
$V_i(e_i)=e_i'X_{i,\rho}e_i$. In the same manner as in~\cite{U6,U7},
by completing the squares, one can establish that 
for all uncertainty signals $\xi_i(t)$,
$\zeta_i(t)$ satisfying the constraints (\ref{Lip}), (\ref{wv.constr}), 
the following dissipation inequality holds 
\begin{eqnarray}
\lefteqn{\mathbf{e}'Q\mathbf{e}+N\Psi(\mathbf{e})+ \dot V(\mathbf{e})\le  -\varepsilon 
V(\mathbf{e})}&& \nonumber \\
&& + \gamma^2\sum_{i=1}^N\left (\|w_0\|^2+\|w_i\|^2+ \sum_{j\in
  \mathbf{V}_i} \|w_{ij}\|^2 \right).
\label{lyap.3}
\end{eqnarray}
The statement of Lemma~\ref{T.aux} now follows from (\ref{lyap.3}). Indeed,
properties (i) and (ii) of Definition~\ref{Def1'} can be established
using the same argument as that used to derive the statement of Theorem~1
in~\cite{U6} from a similar dissipation inequality. Also, it follows from
(\ref{lyap.3}) and the condition $Q>0$ that $\mathbf{e}\in L_2[0,\infty)$. 
This implies that $\dot{\mathbf{e}}\in L_2[0,\infty)$. From the
Cauchy-Schwartz inequality 
$\lim_{t\to\infty}\|\mathbf{e}(t)\|^2=2 \lim_{t\to\infty}\int_0^t
\mathbf{e}(t)'\dot{\mathbf{e}}(t)dt$ exists. Since we have established that 
$\mathbf{e}\in L_2[0,\infty)$, this limit must be equal to 0. Thus,
statement (iii) of Definition~\ref{Def1'} holds as well.  
\hfill$\Box$

\subsection{Proof of Theorem~\ref{main.T}}

Since the matrices $X_{i,\rho}$ are continuous and
piecewise affine, they are differentiable on $\Gamma$ except at $\rho \in
\bar\Gamma$. Using Lemma~6 of \cite{Stilwell-Rugh-2000} and the definition
of $X_{i,\rho}$ in (\ref{X.interp}), it follows that,
given any $\epsilon>0$, there exists a continuously differentiable matrix
function $Y_{i,\rho}$ defined on $\Gamma$, and a constant $\beta>0$ such
that for any `corner' point $c\in\bar\Gamma$,
\begin{gather}
\label{prop-1} \sup_{\rho \in \Gamma}\left\|
Y_{i,\rho}-X_{i,\rho}\right\| < \epsilon, \quad
\rho\in(c-\beta,c+\beta), \\
Y_{i,\rho}=X_{i,\rho}, \quad
\rho\not\in(c-\beta,c+\beta),\\
\sup_{\rho\in \Gamma}\left\| \frac{dY_{i,\rho}}{d\rho}\right\|
\leq \sup_{k}\frac{\left\|X_{i,\rho^{k+1}}-X_{i,\rho^{k}}\right\|}{\bar \rho^k-\underline\rho^{k+1}}.
\label{prop-2.1}
\end{gather}
Note that the approximating matrices $Y_{i,\rho}$ can be chosen symmetric,
  since $X_{i,\rho}$ are symmetric. Also, if a sufficiently small
  $\epsilon>0$ is chosen, positive definite for all
  $\rho\in\Gamma$ matrices $Y_{i,\rho}$ can be selected. 

Now suppose $\rho=\rho(t)\in[\rho^k,\rho^{k+1}]$. Since both
$(\tau_{i,\rho^k}, X_{i,\rho^k})$ and $(\tau_{i,\rho^{k+1}},
X_{i,\rho^{k+1}})$, $i=1,\ldots, N$,  satisfy the LMIs (\ref{T4.LMI.k}),
then due to the linearity of (\ref{T4.LMI.k}),  
$(\tau_{i,\rho},X_{i,\rho})$ are also feasible for the LMIs
(\ref{T4.LMI.k}). 

Let us now consider the synchronization
errors for the system (\ref{plant}), (\ref{nonlin.obs}), and the protocol
(\ref{u}) with the gains defined in (\ref{K.interp}), (\ref{L.interp}). It
is straightforward to verify that the synchronization 
errors satisfy the equation (\ref{enon.fixed.1}) in which $\zeta_i\equiv 0$
and $\rho=\rho(t)$. 
Let us define the vector storage function candidate for this system,
$V_i(e_i,t)=e_i'Y_{i,\rho(t)}e_i$. Also, let 
$\mathbf{w}^i=[w_{ij_1}'~ \ldots~ w_{ij_{p_i}}']'$. 
Since the inequality (\ref{T4.LMI.k}) is strict, the set $\Gamma$ is compact
and $X_{i,\rho}$ is continuous with respect to $\rho$, then it is possible
to choose a sufficiently small $\epsilon>0$ in (\ref{prop-1})
so that the following holds:
\begin{eqnarray*}
\lefteqn{  \frac{d}{dt}V_i(e_i,t)= e_i'\left(\frac{d Y_{i,\rho(t)}}{dt}\right)e_i+
  2e_i'Y_{i,\rho(t)}\dot e_i} && \\
&&
< e_i'\left(\frac{d Y_{i,\rho(t)}}{dt}\right)e_i- e_i'Q_ie_i
-(p_i+q_i)\|e_i\|^2\\
&&+2e_i'\sum_{j\in\mathbf{V}_i}e_j-2\delta_i
V_i(e_i,t) +\sum_{j\in\mathbf{V}_i}\frac{2\delta_j}{q_j+1}V_j(e_j,t) \\
&&
+\gamma^2\|w_0\|^2
+\gamma^2\|w_i\|^2+\gamma^2\sum_{j\in\mathbf{V}_i}\|w_{ij}\|^2.
\end{eqnarray*}

We now observe that for $\rho(t)\in\Gamma\backslash\bar\Gamma$, it follows from
(\ref{prop-2.1}) and (\ref{rate.condition}) that
$ 
\left\|\frac{d Y_{i,\rho(t)}}{dt}\right\|
\le \lambda_{\min}(Q).
$ 
Therefore, 
\begin{eqnarray}
\lefteqn{  \frac{d}{dt}V_i(e_i,t)< 
-(p_i+q_i)\|e_i\|^2+2e_i'\sum_{j\in\mathbf{V}_i}e_j} && \nonumber \\
&& -2\delta_i
V_i(e_i,t) +\sum_{j\in\mathbf{V}_i}\frac{2\delta_j}{q_j+1}V_j(e_j,t)
\nonumber \\
&&+\gamma^2\|w_0\|^2
+\gamma^2\|w_i\|^2+\gamma^2\sum_{j\in\mathbf{V}_i}\|w_{ij}\|^2.
\label{lyap.1}
\end{eqnarray}
The remainder of the proof now follows
using the same argument as that used in the proof of Lemma~\ref{T.aux}.
\hfill$\Box$

\subsection{Proof of Corollary~\ref{Cor}}

The proof is almost identical to the proof of Theorem~\ref{main.T}. The
only difference is to note that the stronger condition
(\ref{rate.condition.strong}) will imply a stronger version of the vector
dissipation inequality (\ref{lyap.1}), with the additional term $-(1-\eta)
e_i'Q_ie_i$ on the right hand side.

%\bibliographystyle{plain} 
%\bibliography{Val,irpnew}

\begin{thebibliography}{10}

\bibitem{GYSS-2012}
H.~F. Grip, T.~Yang, A.~Saberi, and A.~A. Stoorvogel.
\newblock Output synchronization for heterogeneous networks of
  non-introspective agents.
\newblock {\em Automatica}, 48(10):2444--2453, 2012.

\bibitem{HCN-2004}
W.~M. Haddad, V.~Chellaboina, and S.~G. Nersesov.
\newblock Vector dissipativity theory and stability of feedback
  interconnections for large-scale non-linear dynamical systems.
\newblock {\em Int. J. Contr.}, 77(10):907--919, 2004.

\bibitem{LDCH-2010}
Z.~Li, Z.~Duan, G.~Chen, and L.~Huang.
\newblock Consensus of multiagent systems and synchronization of complex
  networks: {A} unified viewpoint.
\newblock {\em IEEE Trans. Circuits Syst.~I: Regular Papers}, 57:213--224,
  2010.

\bibitem{LCCC-2002}
J.~L{\" u}, G.~Chen, D.~Cheng, and S.~{\u C}elikovsk{\'y}.
\newblock Bridge the gap between the {L}orenz system and the {C}hen system.
\newblock {\em International Journal of Bifurcation and Chaos},
  12(12):2917--2926, 2002.

\bibitem{NM-1997}
H.~Nijmeijer and I.M.Y. Mareels.
\newblock An observer looks at synchronization.
\newblock {\em IEEE Trans.~Circuits Syst.~I: Fundamental Theory
  and Applications}, 44(10):882--890, Oct 1997.

\bibitem{OFM-2007}
R.~Olfati-Saber, J.A. Fax, and R.M. Murray.
\newblock Consensus and cooperation in networked multi-agent systems.
\newblock {\em Proceedings of the IEEE}, 95(1):215--233, 2007.

\bibitem{OM-2004}
R.~Olfati-Saber and R.~M. Murray.
\newblock Consensus problems in networks of agents with switching topology and
  time-delays.
\newblock {\em IEEE Trans. Automat. Contr.}, 49:1520--1533, 2004.

\bibitem{PC-1990}
L.~M. Pecora and T.~L. Carroll.
\newblock Synchronization in chaotic systems.
\newblock {\em Phys. Rev. Lett.}, 64:821--824, 1990.

\bibitem{Savkin-2004}
A.V. Savkin.
\newblock Coordinated collective motion of groups of autonomous mobile robots:
  {A}nalysis of {V}icsek's model.
\newblock {\em IEEE Trans. Automat. Contr.}, 49(6):981--982, June 2004.

\bibitem{Shamma-Athans-TAC}
J.~S. Shamma and M.~Athans.
\newblock Analysis of gain scheduled control for nonlinear plants.
\newblock {\em IEEE Trans. Automat. Contr.}, 35(8):898--907, 1990.

\bibitem{SWH-2010}
B.~Shen, Z.~Wang, and Y.~S. Hung.
\newblock Distributed ${H}_\infty$-consensus filtering in sensor networks with
  multiple missing measurements: {T}he finite-horizon case.
\newblock {\em Automatica}, 46(10):1682 -- 1688, 2010.

\bibitem{Stilwell-Rugh-99}
D.~J. Stilwell and W.~J. Rugh.
\newblock Interpolation of observer state feedback controllers for gain
  scheduling.
\newblock {\em IEEE Trans. Automat. Contr.}, 44(6):1225--1229, 1999.

\bibitem{Stilwell-Rugh-2000}
D.~J. Stilwell and W.~J. Rugh.
\newblock Stability preserving interpolation methods for the synthesis of gain
  scheduled controllers.
\newblock {\em Automatica}, 36:665--671, 2000.

\bibitem{Stoustrup-Komareji}
J.~Stoustrup and M.~Komareji.
\newblock A parameterization of observer-ased controllers: {B}umpless transfer
  by covariance interpolation.
\newblock In {\em Proc. American Contr. Conf., 2009.}, pages 1871--1875, 2009.

\bibitem{Swinnerton-Dyer-2001}
P.~Swinnerton-Dyer.
\newblock Bounds for trajectories of the {L}orenz equations: an illustration of
  how to choose {L}iapunov functions.
\newblock {\em Physics Letters A}, 281:161--167, 2001.

\bibitem{TTM-2013}
H.~L. Trentelman, K.~Takaba, and N.~Monshizadeh.
\newblock Robust synchronization of uncertain linear multi-agent systems.
\newblock {\em IEEE Trans. Automat. Contr.}, 58(6):1511--1523, 2013.

\bibitem{U6}
V.~Ugrinovskii.
\newblock Distributed robust filtering with ${H}_\infty$ consensus of
  estimates.
\newblock {\em Automatica}, 47(1):1 -- 13, 2011.

\bibitem{U8a}
V.~Ugrinovskii.
\newblock Gain-scheduled synchronization of uncertain parameter varying systems
  via relative ${H}_\infty$ consensus.
\newblock In {\em Proc. Joint 50th IEEE CDC and ECC}, pages 4251--4256, 2011.

\bibitem{U7}
V.~Ugrinovskii.
\newblock Distributed robust estimation over randomly switching networks using
  ${H}_\infty$ consensus.
\newblock {\em Automatica}, 49(1):160--168, 2013.

\bibitem{LaU1}
V.~Ugrinovskii and C.~Langbort.
\newblock Distributed ${H}_\infty$ consensus-based estimation of uncertain
  systems via dissipativity theory.
\newblock {\em IET Control Theory \& App.}, 5(12):1458--1469, 2011.

\bibitem{YSS-2011}
T.~Yang, A.A. Stoorvogel, and A.~Saberi.
\newblock Consensus for multi-agent systems --- synchronization and regulation
  for complex networks.
\newblock In {\em Proc. American Contr. Conf., 2011}, pages 5312--5317, 2011.

\bibitem{YUMP1}
M.~Yoon, V.~Ugrinovskii, and M.~Pszczel.
\newblock Gain-scheduling of minimax optimal state-feedback controllers for
  uncertain linear parameter-varying systems.
\newblock {\em IEEE Trans. Automat. Contr.}, 52(2):311--317, 2007.

\end{thebibliography}

\newcommand{\noopsort}[1]{} \newcommand{\printfirst}[2]{#1}
  \newcommand{\singleletter}[1]{#1} \newcommand{\switchargs}[2]{#2#1}

\end{document}